# Tracking and controlling monolayer water in gold nanogaps using extreme plasmonic spectroscopy

Elle W. Wyatt[1], Sarah May Sibug-Torres[1], Rakesh Arul[1], Marika Niihori[1], Tabitha Jones[1], James W. Beattie[1], Bart de Nijs[2], Jeremy J. Baumberg[1*]

[1] NanoPhotonics Centre, Cavendish Laboratory, Department of Physics, JJ Thompson Avenue, University of Cambridge, Cambridge, CB3 0HE, United Kingdom
[2] Physics for Sustainable Chemistry Group, Department of Physics, JJ Thompson Avenue, University of Cambridge, Cambridge, CB3 0HE, United Kingdom

*Corresponding author: Jeremy J. Baumberg
Email: jjb12@cam.ac.uk

## ABSTRACT
Nanogaps are ubiquitous across science, confining molecules and thus changing chemistries which influence many areas such as catalysis, corrosion, photochemistry, and sensing. However in ambient conditions, it is unclear how water solvates nanogaps and even if nominally dry, what water structure persists. Despite its low Raman cross-section, surface-enhanced Raman spectroscopy (SERS) enables study of water at coinage metal surfaces. Using multi-layer aggregates of close-packed gold nanoparticles with sub-nanometre gaps precisely defined by organic spacer molecules, we achieve consistent and large SERS enhancements, enabling systematic study of water within these confined spaces. Ostensibly dry facets in air evidence water monolayer coatings, with hydrogen-bonding only reappearing upon immersion in solution. Under negative applied potentials, surface water is seen to re-orient at the metal facets with distinct spectral shifts among the quartet of vibrational peaks which correspond to those expected from water dimers. Comparing nanogaps in deuterated water also reveals how individual water molecules bind onto organic spacer molecules in such nanogaps. Realistic models of water dressing will enable better understanding of catalytic and contact chemistries.

## INTRODUCTION
Surface Enhanced Raman Spectroscopy (SERS) is a powerful technique for studying molecules adsorbed on metal surfaces, offering both information about the structure of the molecules and their interaction with the surface, even at low concentrations[1]. It is particularly suited for in situ and in operando measurement compared to other surface sensitive techniques. However, the Raman spectrum of water is broad and weak making it difficult to detect, especially when present as only a few monolayers on a surface. Using SERS to detect water molecules at metal nanostructure surfaces enables the study of interfacial and confined water, which appears in many natural systems and has implications for catalysis, corrosion, photochemistry, molecular electronics, and sensing[2,3]. The ability to study surface-water interactions in situ using simple CW spectroscopy, compared to complex ultrafast experiments of scanning probe microscopes, would be extremely valuable[4].

Spectroscopic and simulation studies demonstrate that structuring occurs when water is present on a surface or in a pore. Changes in the Raman spectra of water molecules indicate changes in their hydrogen (H-)bonding networks, with a decrease in H-bonding seen for more confined water molecules[5–11]. Previous SERS studies of surface water have employed roughened silver, gold, or platinum electrodes and aimed to study water hydrogen bonding networks when potentials are applied to the metal[7,12–14]. Even with SERS, many of these studies suffer from low signal-to-noise due to the low water Raman cross section and require an applied negative potential in the presence of highly concentrated (>0.1 M) electrolyte to make the water peaks appear, potentially due to co-adsorption of water and ions at the metal surface under applied potential[15,16]. Studies of surface water when not in such electrolytes are rarely performed at ambient conditions, but instead at low temperatures and under vacuum where isolated clusters of water molecules can be observed[17–22].

Here the use of extreme optical confinement within precision nanogaps allows us to comprehensively study the behaviour of water molecules at metal surfaces under a wider variety of conditions. Recently we demonstrated the use of self-assembly to fabricate a reproducible SERS substrate consisting of a near-monolayer aggregate film of close-packed gold nanoparticles ('MLagg')[23–25]. The gaps between nanoparticles are precisely defined by rigid cucurbit[n]uril spacer molecules (CB[n], n=5-8), giving reproducible 0.9 nm interparticle distances that provide large SERS enhancements for molecules in the plasmonic hotspots created between the nanoparticles. CB[n] is known to encapsulate between 2 and 12 water molecules, depending on the CB[n] size[26–28], previously used to study the effects of confined water inside the CB[n] cavity[29]. Here, we are now able to study the water present outside CB[n], in nominally dry nanostructures when in air, and demonstrate that different (hysteretic) water structures arise for MLaggs in water and upon drying. We compare this with immersion in dimethyl sulfoxide (DMSO) which strips all water from the nanogaps. Using electrochemical (EC-) SERS we track water re-orientation on the nanoparticle surface at negative voltage, going beyond previous studies to distinguish the behaviour of individual water SERS peaks and its interaction with organic molecules.

## RESULTS AND DISCUSSION
### Water in nanogaps

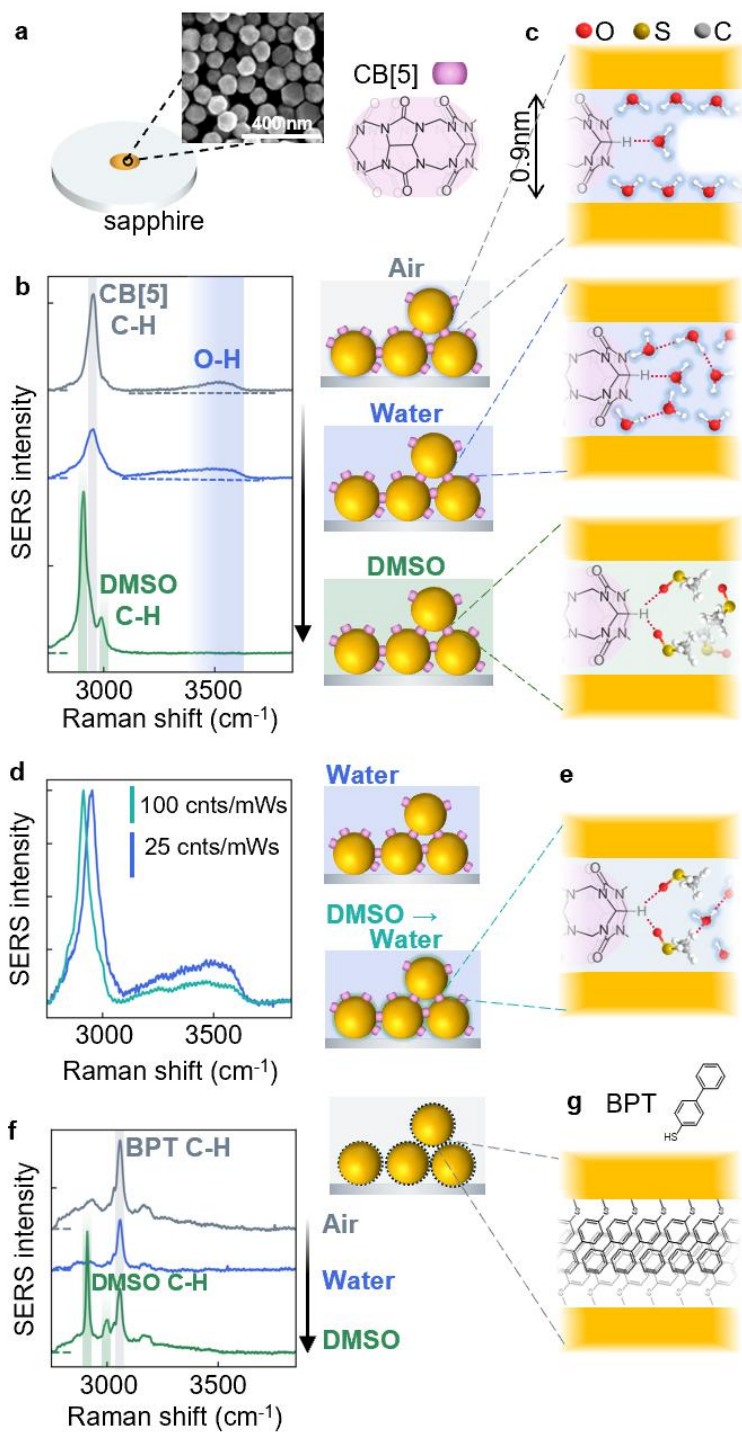

**Figure 1. MLagg SERS in air, water, and DMSO.** (**a**) Self assembled AuNP sheet immobilized on sapphire substrate, with nanogaps defined by CB[5] molecules. Inset shows SEM image of MLagg with 80 nm AuNPs. (**b,c**) SERS spectra in air, water, and DMSO, with sketches indicating the changes in water hydrogen-bonding network (red dashed) and exclusion by DMSO. (**d,e**) DMSO persists in MLagg nanogaps even after drying

and re-immersion in water (green), yielding permanent changes in O-H and CB[5] C-H peaks compared to direct immersion in water (blue). (**f,g**) BPT excludes all water from the nanogaps.

Previous work has demonstrated MLaggs to be a reproducible and sensitive SERS substrate for sensing a wide range of analytes[23–25]. MLaggs consist of ~1 monolayer of AuNPs that are aggregated at a liquid-liquid interface using cucurbit[5]uril (CB[5]) and deposited on a sapphire substrate to minimise fluorescence backgrounds in high wavenumber SERS measurements (see Methods). This structure has been well characterised using SEM (Figure 1a) and AFM, while TEM confirms the 0.9nm nanogaps (Figure S1, S2, Supplementary Note 1)[23–25]. Pristine nanogaps are generated by oxygen plasma cleaning the MLagg which removes all ligands and oxidises the top three gold atomic layers and then regenerating the nanogaps ('re-scaffolding') by reintroducing a scaffolding molecule into the gaps whilst chemically reducing the atomic Au surface (Figure 1a)[23]. The high (>$10^6$) signal enhancement in the nanogaps[25] enables greatly improved signal-to-noise of the SERS water spectrum compared to all previous studies. This opens the opportunity to study water in precise and uniform sub-nm gaps between nanoparticles, unachievable using roughened metal surfaces. The cleaning and regeneration step is crucial to provide reproducible water spectra from gold facets with well-defined surface functionalisation and observe surface water even in air.

After re-scaffolding with CB[5], rinsing in DI water and drying in $N_2$ flow, the MLagg SERS shows a broad water band ~3500 cm$^{-1}$ (Figure 1b). This can be attributed to the O-H stretching mode of water molecules within a restricted H-bonding network, where most molecules remain without H-bonds, as expected for surface or nanogap water layers[5,6,8,30]. We note that the AuNP surface is hydrophilic due to the presence of CB[5] and Cl$^-$ after re-scaffolding, so this is slightly redshifted from the position expected for a hydrophobic surface. The CB[5] C-H stretching mode is also evident as a sharp peak at 2955 cm$^{-1}$, with contributions from the CB[n] equatorial CH and CH$_2$ (DFT in Figure S8). After immersing the sample in DI water for 30 minutes, the O-H peak is broadened to lower wavenumber compared to in air, and the CB[5] C-H peak also broadens, but does not shift. This is indicative of an increased water H-bonding network causing a redshift in O-H peaks as electron density is withdrawn from the O-H bond[6,31]. MLaggs have previously been shown to be robust to wetting and are stable in both liquid and gas flow[23–25], remaining fixed to substrates during immersion and drying.

When the sample is then dried and placed in DMSO for a further 30 minutes, all water is stripped from the Au surfaces and the O-H stretch SERS peak disappears (Figure 1b). Two DMSO C-H stretch peaks appear at 2905 cm$^{-1}$ and 2990 cm$^{-1}$ and the CB[5] C-H peak is redshifted compared to in air and water. In the low wavenumber region of the spectrum (Figure S3a), the characteristic CB[5] peaks at 450, 830 cm$^{-1}$ decrease because they are partially replaced by DMSO at the nanogap facets. This results from the high affinity of DMSO to the Au surface, which allows it to competitively bind (suggested by removal of some CB[5]), as well as its capability to solvate individual water molecules thus helping their removal from the Au surface[32].

These spectra give insight into the structure of water molecules present in the nanogaps (Figure 1c). Water in nanogaps has been modelled, but not studied extensively experimentally, with most SERS studies of water performed on heterogeneous roughened metal surfaces[5,6,33]. For gaps <1 nm in width, it has not been clear how solvation occurs, although this can be crucial for many applications such as catalysis and gas sensing where solvation of the surfaces plays an important role[2,3]. In air, our data shows only the higher wavenumber O-H peaks (Figure 1b; grey, ~3500 cm$^{-1}$), evidence of a suppressed H-bonding network. This implies that water molecules form a monolayer (or a partial monolayer surrounding the CB[5] molecules) on the Au surfaces without mid-gap (dimer) water molecules. The larger interstices between three or more nanoparticles are too large to be fully filled by capillary condensation. We note that while the water monolayer likely coats the entire AuNP surface, only the SERS hotspots inside the nanogaps are probed here. When additional water is reintroduced, extra water molecules move into the nanogaps, increasing the number of hydrogen bonds per water molecule and new lower wavenumber O-H peaks appear (Figure 1b; blue, ~3250 cm$^{-1}$). In DMSO, it is evident that all water molecules in the gap are replaced (as all the O-H peaks disappear), while it also interacts with the CB[5] C-H bonds (shifting them to lower wavenumber, Figure 1b; green).

Part of this DMSO coating is found to persist in the nanogaps. Even after drying and immersing the MLaggs in DI water, >10% of the low wavenumber DMSO SERS peak remains (C-S line at 670 cm$^{-1}$, Figure S3b,c). Additionally, an irreversible redshift of 80 cm$^{-1}$ is observed for the CB[5] C-H peak after DMSO introduction (2915 cm$^{-1}$, Figure 1d). The water O-H peak for DMSO treated MLaggs is also ~50% weaker compared to substrates directly immersed in water. This suggests >10% nanogap DMSO remains strongly bound to the CB[5] and the Au surfaces (Figure 1e, S3b), with solvated water returning into the middle of the gap.

Similar effects are seen for MLaggs re-scaffolded with the larger diameter CB[6-8] (Figure S4a) for the same measurements in air, water, and DMSO. The ratio of the integrated O-H peak to the CB[n] C-H peaks (when normalised by the CB[n] portal area, spacing on the Au surface, and polarizability), is proportional to the CB[n] circular area (Figure S4, Supplementary Note 2). Since this estimates the number of $H_2O$ to CB[n], it suggests that dense structures of larger CB[n]s accommodate more surrounding water molecules per CB[n].

Re-scaffolding the gaps with a thiol such as biphenyl-4-thiol (BPT) fully excludes water from the Au surfaces. No SERS water peak is seen either in air or in water (Figure 1f), which is attributed to the dense packing and hydrophobic nature of BPT (Figure 1g). In DMSO, similar DMSO C-H peak positions are seen compared to the CB[5] MLaggs, but no effect on the BPT C-H stretch is observed. This shows how changing the scaffolding ligand in the nanogaps changes their hydrophobicity and interactions with other molecules.

## Scaffolding nanogaps in deuterated solutions

In deuterated solutions, H-bonding is weaker than in water. Plasma-cleaned MLaggs are thus re-scaffolded in either aqueous or deuterated solutions of CB[5] and NaCl at low [HCl] (see Methods). The deuterated CB[5] solutions require ~1 month incubation to allow H-D exchange of water molecules in the CB[5] cavities[29]. There is a clear difference between the O-H and O-D stretch peaks as expected from the isotope effect (Figure 2a), with the O-D peak shifted to lower wavenumber by ~1000 cm$^{-1}$. The peaks are narrower and enhanced, as generally observed in deuterated water spectra, arising from slower dynamic motion of the $D_2O$ molecules and changes in intermolecular coupling strength[34–37]. All CB[5] peaks are also slightly shifted when in $D_2O$, as expected (Figure S6a).

After iteratively fitting and subtracting a 2$^{nd}$ order polynomial background (see Methods), the SERS spectra obtained in this system show four peaks which correspond to the water dimer Raman modes previously reported [21,22,29,31] (Figure 2b, details of fitting protocol in Supplementary Note 3). This results from splitting of each of the antisymmetric (AS) and symmetric (SS) O-H or O-D stretches into two peaks from the hydrogen-bonding perturbation between two molecules. The O-H stretches of the H-bond donor molecule('d') are shifted to lower energies because the additional hydrogen bonding to the acceptor ('a') withdraws electron density from the covalent bonds (Figure 2c). Comparing the spectrum in air (grey line Figure 2c) shows the two lower wavenumber O-H peaks only appear when the sample is wet, implying that dimer modes are only present in wet samples when additional water fills the nanogaps. For Raman of bulk $H_2O$ or $D_2O$, the higher wavenumber peaks corresponding to surface water are instead suppressed (Figure S5a). The lower O-H or O-D peaks are seen to be broader, likely because there are more dimer configurations, but the sharper $D_2O$ spectra clearly constrains the peak fits allowing their subsequent independent tracking.

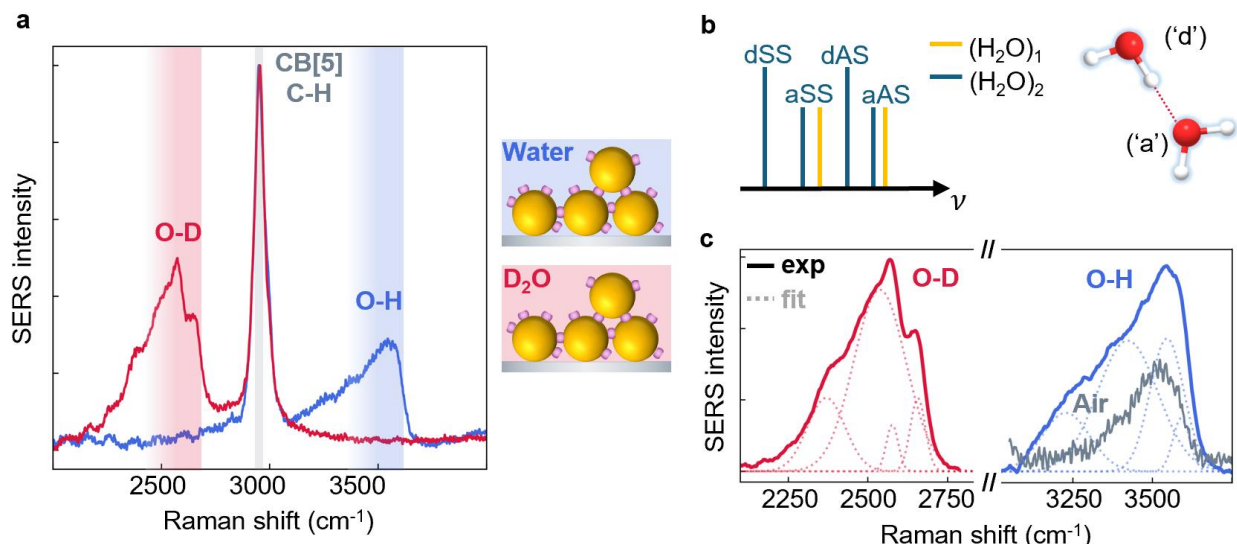

**Figure 2. MLaggs re-scaffolded in $H_2O$ and $D_2O$.** (**a**) MLaggs scaffolded in aqueous or deuterated CB[5] solutions show O-D (~2500 cm⁻¹) or O-H peaks (~3500 cm⁻¹). Spectra normalized to C-H peak. (**b**) Vibrational energy splitting for dimer (blue, single $H_2O$ in orange) of antisymmetric (AS) and symmetric (SS) O-H peaks when H-bonding is present. Schematic to right shows donor ('d') and acceptor ('a') molecules. (**c**) Water peaks, fit with four gaussians (dotted lines). Spectrum in air (grey) shows absence of two lower wavenumber peaks when dry.

Repeating this experiment for CB[7] scaffolded MLaggs gives identical results to CB[5] (Figure S6b), where the O-H and O-D peaks fit to the same four peaks (Figure S6c) with the same positions and peak ratios (Figure S6d). This suggests that after plasma cleaning, most water is outside CB[n] cavities since the peaks from water inside the cavities change with CB[n] size[29].

## SERS of drying nanogaps

We now explore how nanogaps dry after water immersion. After removal from the aqueous or deuterated scaffolding solution the CB[5] samples are placed in a 50 sccm $N_2$ flow and SERS measured continuously for 30 minutes at equal time intervals (Methods). Changes in the SERS peaks over time can be seen in both samples (overlaid in Figure 3a). For the sample scaffolded in $H_2O$, there is a slight decay of the O-H peak over the first 200 s (Figure 3b, blue curves). In contrast, in the $D_2O$ sample (red curves), the decay of O-D occurs abruptly within 200 s and is replaced by O-H over similar timescales. Some water remains in the nanoparticle gaps after drying of both samples, bound as a monolayer to the facets as before.

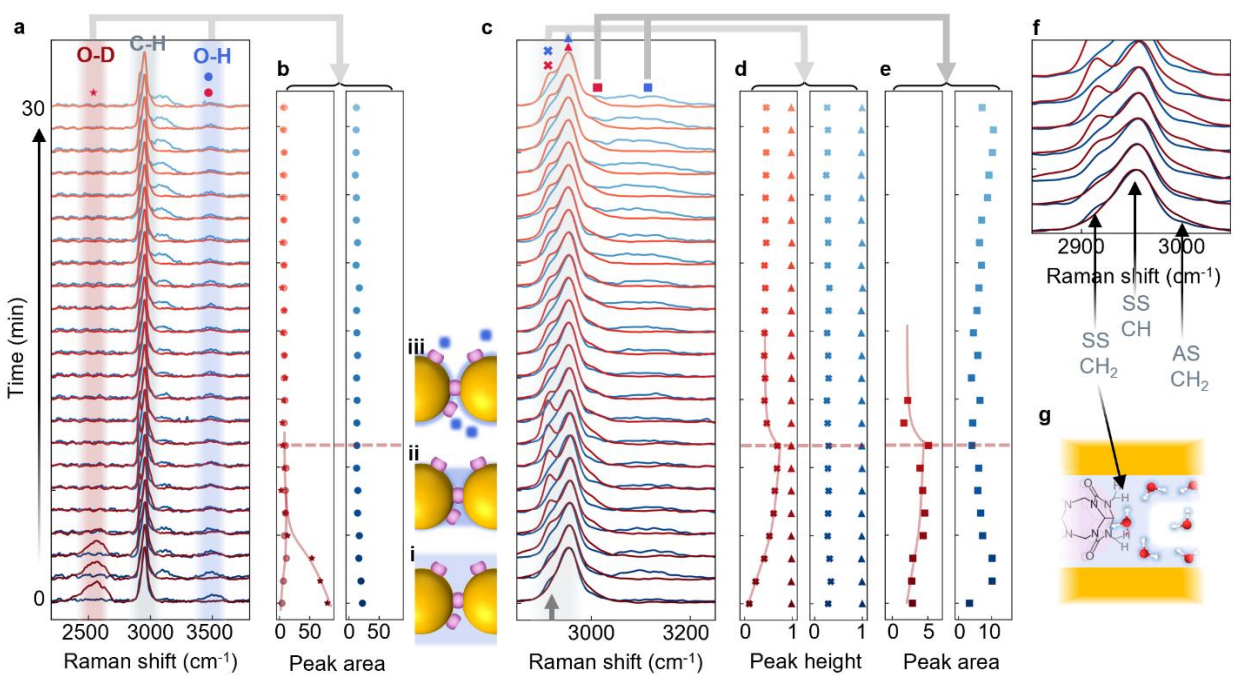

**Figure 3. Drying of MLaggs after re-scaffolding in $H_2O$ or $D_2O$.** (**a**) MLaggs are dried in $N_2$ flow and SERS measured at equal time intervals for 30 minutes. (**b**) For $D_2O$, the O-D line decays abruptly within 200 s while weaker O-H line rises. For $H_2O$, O-H peak slightly decays over first 200 s. **i-iii** depict water in nanogap. (**c**) Expanded plot of C-H peaks from CB[5] shows changes when drying. For $D_2O$ sample, a second sharp C-H peak grows rapidly at 2920 cm$^{-1}$ until 500 s (red dashed line), peak heights extracted in (**d**). (**e**) Area of C-H peaks emerging at ~3000 cm$^{-1}$ and ~3100 cm$^{-1}$ in the two samples. (**f,g**) Attribution of C-H lines from DFT, to the CH and $CH_2$ stretches of CB[5], and sketch of possible structure when perturbed on drying from $D_2O$ (**g**).

However, in this system another timescale appears in the C-H lines of CB[5]. Drying from $H_2O$ the ratio of C-H lines remains constant, but in the $D_2O$ sample the C-H line at 2915 cm$^{-1}$ (grey arrow) becomes intense, reaching its maximum ~500 s (Figure 3c,d). Over 30 mins, despite weakening again by 40%, this line remains persistently stronger than for the $H_2O$ sample. Over the same timescale, a slight shift (~5 cm$^{-1}$) is also seen in the 2955 cm$^{-1}$ C-H peak (Figure S7a).

From DFT of CB[5] (Figure S8a), the C-H lines can be attributed to the CH and $CH_2$ stretches (Figure 3f,g). DFT predicts four peaks (symmetric and antisymmetric stretches for each type of C-H bond), with three dominating the Raman spectrum (Figure S8b). All changes in the C-H lines occur over the same timescale of 500 s, so this implies water interacts with the N-$CH_2$- group on the CB[5] (Figure 3g)[38]. It appears that when the interaction with $D_2O$ occurs in this position, it greatly increases the net Raman cross section of the CB[5] $CH_2$ stretch. The slow escape suggests that it is also bonded to the Au, and thus has a high desolvation energy barrier. Further DFT or modelling will be crucial to understand this interaction, but as we suspect the Au surface plays a key role, achieving optimised structures and Raman spectra is challenging. We highlight this shows the capability of SERS to track detailed solvation of molecules.

In summary, the solvated $D_2O$ rapidly escapes from the nanogap hotspot, but some $D_2O$ bind more slowly to the CB[5] and then remains for surprisingly long times. The two timescales could arise from different water diffusion and evaporation rates from the bulk and the nanogaps (Figure 3bi-iii). The nanogaps are initially immersed completely in water (i), with rapid bulk evaporation removing most $D_2O$ (ii). This leaves remaining $D_2O$ in the nanogaps to bind to the CB[5], and which escapes much more slowly (>30 minutes) due to its high binding association (iii).

In both samples a higher energy line also appears (Figure 3c,e, S7b), at 3100 cm$^{-1}$ for the $H_2O$ sample, and at 3000 cm$^{-1}$ for the $D_2O$ sample, which rises over a similar timescale to the 2955 cm$^{-1}$ line seen for the $D_2O$ sample, suggesting they are related. Comparing this to the DFT of CB[5] in implicit water (Figure S8a), a C-H line at 3100 cm$^{-1}$ can be seen which has not been previously observed in the SERS spectra. It suggests that on drying, this C-H line is enhanced in a similar way to the 2955 cm$^{-1}$ line. Alternatively, SERS peaks for ice-like water appear in this region. In either case, a strong interaction between the $D_2O$ or $H_2O$ and the CB[5] is indicated. This effect is only possible to see when comparing the samples drying from $D_2O$ and $H_2O$, but is currently not understood as when drying from $H_2O$ it persists for >30 min.

Heating a CB[5] scaffolded sample from room temperature (22$^o$C) to 80$^o$C in 50 sccm $N_2$ flow (Figure S9a), shows a systematic decrease in area of the O-H lines by 90%, which is recovered on allowing the sample to cool once more (Figure S9b). At the same time the C-H area only slightly decreases on heating. This suggests that as the water molecules are thermally driven off, there is some re-organisation of the CB[5] molecules and Cl$^-$ ions (remaining from the re-scaffolding step, ~5 Cl$^-$ ions per CB[5] molecule (see Figure S10 and Supplementary Note 4) on the gold (Figure S9c). Using the change in O-H SERS peak area to quantify changes in surface coverage, the enthalpy and entropy of binding can be estimated as $\Delta H \sim 30$ kJ/mol (300 meV) and $T\Delta S = 25$ kJ/mol at 300K (Supplementary Note 5). This is comparable to other calculated binding energies of water[39–41], while their similar value leads to the large observed changes in water coverage near room temperature.

## EC-SERS of water in MLaggs

The MLagg system also offers the opportunity to study effects of applying potential to the metal facets to investigate surface binding of molecules[25]. Previous studies on surface water while applying a potential have been limited to roughened silver, gold or platinum electrodes, which have a heterogeneous electrical double layer and thus variable local field at the surface due to widely different hotspot sizes and thus variations in applied potential. In contrast, MLaggs have precisely defined hotspot width from the CB[5] scaffold, and thus have more controlled potentials in all gaps. This is vital to see the four distinct water peaks observed (Figure 2), which we now track under applied potential. The MLaggs are placed in an EC cell (in pH7 1 M phosphate buffer, Methods) and potential is applied to the MLagg for 30 s in 0.1 V steps from 0 to -0.8 V (*vs* Ag/AgCl) and back, while the SERS spectra are measured (Figure 4a). Either water or deuterated buffer solution is used. In both cases, the O-H or O-D lines grow and shift to lower wavenumber as the potential becomes more negative (Figure 4b,c).

This is consistent with previous electrochemical studies performed on water at roughened silver or gold electrodes, where broad SERS water lines behave similarly[7,12,13]. This is understood as reorientation of water at the electrode surface. In some studies, the water molecules are thought to lie flat with respect to the surface for ambient conditions (at 0 V). However, due to the SERS selection rules, bonds lying in such an orientation would have negligible Raman cross section, while bonds perpendicular to the surface (and parallel to the optical field in the nanogaps) have the strongest cross section[42]. As we see O-H peaks at all applied potentials and in air, and the peak intensity does not decrease during the potential sweep, the water molecules cannot be lying flat on the Au surface at any potential[43]. At open-circuit potential (OCP), which is slightly positive (around +0.1 to +0.2 V), the molecules are configured with oxygen atoms down. When the potential is decreased first to 0 V some initial restructuring may occur (but with no loss in SERS intensity), and at more negative potentials the water molecules re-orientate so the hydrogens face the surface (Figure 4d,e). This forms O-H--Au bonds and charge transfers from the metal into the antibonding orbital, weakening the O-H bond and thus redshifting the SERS peak. This also disrupts the hydrogen bonding network of water as these molecules can no longer act as hydrogen bond donors[7,11,16,44,45].

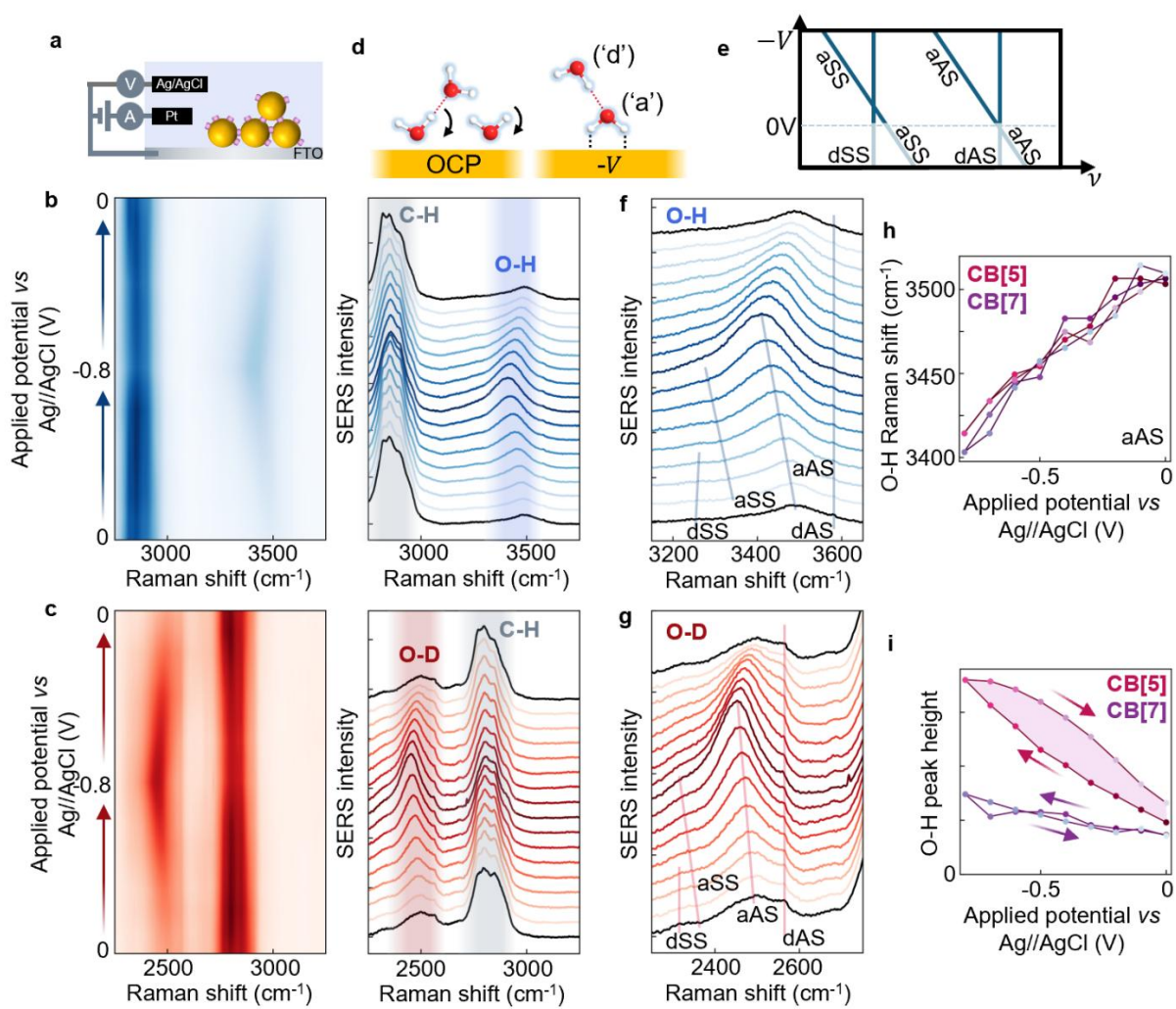

**Figure 4. EC-SERS for CB[5] MLaggs in 1M phosphate buffer.** (**a**) Electrochemical cell set-up. (**b,c**) SERS spectra for CB[5] MLaggs applying potentials from 0 to -0.8 V (*vs* Ag/AgCl) and back in 0.1 V steps, (**b**) in 1 M phosphate buffer, and (**c**) in 1 M deuterated phosphate buffer. Main O-H and O-D peaks strengthen and shift to lower wavenumber at negative potentials. (**d,e**) Reorientation of water on gold surface at negative surface potential, (right) corresponding to shifts in the acceptor peaks (a) in the dimer model. (**f,g**) Expanded plot of O-H and O-D peaks, with shifts of four dimer peaks indicated (lines). (**h,i**) O-H aAS peak shifts and amplitudes for both CB[5] and CB[7] samples, hysteresis for CB[5] shaded.

In contrast to previous electrochemical studies, here we clearly track the four distinct peaks that make up the O-H and O-D bands (Figure 4f,g). In each band, two lines shift with potential while two remain static. Comparing expectations from this (simplistic) dimer model (Figure 4e),[29] the donor water being remote from the Au thus experiences no charge transfer (and thus no shifts). On the other hand, the acceptor water is hydrogen down on the Au, enabling significant charge transfer which lowers the electron density in the O-H bond, thus weakening it. This causes redshifts in the acceptor symmetric (SS) and antisymmetric (AS) lines, redshifting them by ~80 cm$^{-1}$ [20]. Both acceptor lines become >300% stronger by -0.8 V, dominating the overall O-H band. We suggest this may arise from increasingly oriented molecules at negative potential or increasing polarisability of the bond due to the charge transfer. Only small changes in intensity are seen in the low wavenumber bands of CB[5] (Figure S11) and in the CB[5] C-H lines, indicating that there is no loss or re-arrangement of CB[5] on the surface.

These effects are similarly consistent for CB[7] (Figure S12a), however for BPT scaffolded MLaggs (Figure S12b,c), exclusion of all H$_2$O and D$_2$O from the nanogap results in no changes in either peak intensity or position. This shows the thiols form an impermeable layer, preventing water being brought down to the surface, even at negative potentials. Comparing peak shifts of the O-H aAS line for CB[5] and CB[7] samples gives identical results (Figure 4h) as expected if they only play a role in scaffolding the gaps. However the peak heights are found to be scaffold dependent (Figure 4i), and hysteresis appears for CB[5]. This suggests some effect of the electrical double layer and local ions.

The dependence of these effects is also tested in different phosphate buffer concentrations of 100 mM and 1 mM (Figure S12d). Little effect is seen on the peak shifts, but the O-H peak height increases slightly as the buffer concentration increases (20% for a x100 buffer concentration increase)[13]. In addition, at higher buffer concentrations, increased hysteresis is observed, indicating that the changing double layer thickness may influence the rotation of the water molecules close to the Au facet surfaces. We note that the Debye length is larger than the gap size for the lowest buffer concentration, but much smaller than the gap at the highest concentration, supporting this suggestion. We also note that most previous work does not examine cyclic scans, and thus hysteresis does not seem to have been previously observed.

Additional metastability is observed, related to surface charge. The O-H and C-H lines are different with the sample in the EC cell compared to the previous solution conditions (Figure 1-3). Even at 0 V, the H$_2$O and D$_2$O lines already are slightly redshifted compared to other conditions (Figure S13). In the EC cell, the potential is calibrated *vs* a Ag/AgCl reference which is difficult to directly compare directly to these other conditions. Previous studies have shown peak shifts to below 3500 cm$^{-1}$ can occur in electrolytes and at negative potential (but only shown for < -0.3 V and not for 0 V)[13,16,43]. We suggest that the surface charge of the MLagg at OCP is slightly positive (around +0.1 to +0.2 V), bringing the water O atom closer to the gold. Application of 0 V can give thus an initial change in water orientation, as this constitutes a slight displacement from the ground state of the MLaggs in the EC cell. A second effect can be the high ion concentration in the buffer. Previous studies have reported a decrease in the linewidths and changes in the water peak ratios, caused by a disruption in the hydrogen bonding network of water by ions[13,46], but depending on the ions present, the peaks either blue- or redshift[13]. To check for the effects of ion concentration, the HCl concentration in the MLagg system was altered (Figure S14a). As the HCl concentration increases from 0 to 5 M, the low wavenumber Au-Cl peak at 250 cm$^{-1}$ initially decreases (1 M) but then increases with further salt, while the O-H peak ratios indeed become more surface-water-like (Figure S14b,c).

We note also additional changes due to surface charge. A significant shift and broadening is seen for the C-H lines in the EC cell (Figure S15a): a narrow ($\alpha$) peak is observed in the wet and dry states, but when in the EC cell the C-H line consistently broadens as lower ($\beta$) peaks appear in the SERS spectrum. Small variations in the C-H line ratios are also seen with changing voltage. To identify the origin of this effect two identical CB[5] MLagg samples are immersed in water and then in 0.5 M NaCl. One is placed in a cell with no electrodes, and the other in the EC cell at OCP (Figure S15b,c). The $\beta$ C-H peaks only appear with NaCl in the EC cell, so it appears that induced currents when counter-electrodes are present modify the nanogap organisation. The solvent history of nanogaps can thus be important in determining the state of organics and water in these confined geometries. This opens up new areas of exploration that can explain many metastable surface conditions that have proved puzzling.

While we have presented here simple models based on dimer donor-acceptor H-bonding, we ignore the full complexity of interactions. Current DFT and Monte-Carlo models find it challenging to capture the effects going on here (with nanogaps, ions, and fields), although they can suggest lowest energy states for the molecular organisation (however only without applied potentials, and ions)[47–53]. Without vibrational data from such simulations, it is not possible to carefully corroborate with the detailed results above. This system is thus a prototypical dataset for advancing theoretical models of water at surfaces, and should spur progress.

## CONCLUSION

Using precision gold nanostructures, we show that in ambient conditions monolayers of water are trapped in sub-nm nanogaps, and can only be removed by solvents with a stronger affinity for Au such as DMSO and thiols, or by heating. With these MLagg SERS substrates it is now possible to watch the interaction between water molecules and a metal surface under a range of conditions including varying temperatures, salt concentrations, and applied electric potentials. We evidence clear interactions of water with organic molecules in the nanogaps, and discover that deuterated solutions show significant and unexpected differences with their water solvation, which can be long lasting. We find a 4 band vibrational model based on H-bonded water dimers provides a reasonable account of the data, but that full quantum chemical models required are not yet available to simulate even this restricted system. The ability to study a wide range of water solvation conditions at metal surfaces with bound reactants and analytes should open a diverse landscape of science and technology applications.

## METHODS

### Monolayer aggregate preparation.

500 µL each of chloroform ($CHCl_3$) and commercial (BBI Solutions) citrate-capped 80 nm gold nanoparticles (AuNPs) were added to an Eppendorf. 100 µL of 1 mM cucurbit[5]uril (CB[5]) solution was then added and shaken for ~1 min to initiate aggregation. The mixture was left to settle for the immiscible $CHCl_3$ and aqueous phases to separate and the aggregated AuNPs to move to the phase interfaces (chloroform-aqueous and aqueous-air). The aqueous phase was washed with three 300 µL aliquots of DI water to dilute the citrate salts and other supernatants, then concentrated by careful removal of the aqueous phase to form a ~5 µL aggregate droplet floating on the $CHCl_3$. The droplet was deposited onto a pre-cleaned sapphire coverslip (0.15 mm thick) to minimise background fluorescence, or an FTO glass slide for EC-SERS measurements. Once dried, the resulting AuNP multilayer aggregate (MLagg) was rinsed with DI water and dried with $N_2$.

The MLaggs were oxygen plasma cleaned for 45 minutes (oxygen mass flow of 30 sccm, 90% RF power) using a plasma etcher (Diener electronic GmbH & Co. KG) to remove CB[5], citrate and other supernatants from the AuNP surfaces (verified using SERS). To re-introduce a scaffolding ligand, the MLaggs were either immersed in 1 mM CB[5] or CB[7] solution prepared in 1 M HCl for 5 minutes, or in 10 mM BPT in ethanol for 30 minutes. In both cases, the MLaggs were rinsed with DI water and dried with $N_2$.

To compare $H_2O$ and $D_2O$ re-scaffolding, 1mM solutions of CB[5] and CB[7] were prepared in $D_2O$ and left for ~1 month for H-D exchange to take place. Plasma cleaned MLaggs were immersed in 2 µL 1 M HCl with 198 µL 1 M NaCl and 200 µL 1 mM CB[5] or CB[7] solutions in DI water or equivalent deuterated solution for 10 minutes.

All chemicals were purchased from Sigma-Aldrich and aqueous solutions were prepared in deionized water (>18.2 MΩ cm$^{-1}$, Purelab Ultra Scientific system).

### SERS measurements.

SERS measurements were collected on a Renishaw InViva Raman confocal microscope, using a 20x objective (NA = 0.4) and 1200 lines mm$^{-1}$ grating. To collect the lower wavenumber spectral regions, the 785 nm excitation laser was used in extended scan mode, with 1 s integration time and 0.5% laser power (2.2 mW at the sample). To combat instrument sensitivity drop off, a 633 nm laser was used to collect SERS spectra in the region 2000 - 4000 $cm^{-1}$, with either 10 s or 30 s integration time and 10% laser power (0.37 mW at the sample). All measurements were taken at room temperature and the spectra were calibrated with respect to Si. Spectra were analysed using python.

For high wavenumber data, the raw spectral data were smoothed, and a $2^{nd}$ order polynomial background was iteratively fit to the baseline of the peaks. Each spectrum was checked to ensure that the background was not being overfitted.

### Drying experiments.

A simple flow cell was made by laser etching (Needham Laser Tech N-Lase Desktop Pro) a well (diameter 5 mm) and an input and output line into a 5 mm thick stainless steel block. Peek tubing with internal diameter 1/32″ was glued into the inlet and outlet channels. The input was connected to a pure $N_2$ flow controlled by a mass flow meter (Alicat MC-Series) and the output to an exhaust line. A washer of Teflon tape was placed over the well of the flow cell. MLaggs were removed from the re-scaffolding solution and placed without drying face down in the centre of the well. The MLagg was pressed down and taped in place. As the SERS timescan was started, the $N_2$ flow was set to 50 sccm.

### EC-SERS measurements.

An EC-SERS flow cell was machined in Teflon to incorporate a standard three-electrode electrochemical system whilst allowing optical access to the MLagg through a glass cover slip. The three-electrodes were made up of a leakless Ag/AgCl reference electrode (LF-1-45 from Innovative Instruments Ltd), a Pt wire (Sigma-Aldrich) counter electrode, and the working electrode was the MLagg SERS substrate on FTO-coated glass. The cell accommodates a thin 0.5 mm layer of electrolyte between the MLagg working electrode and the glass coverslip. Electrolyte solutions (0.01, 0.1 and 1 M potassium phosphate in DI water or $D_2O$) were inserted using a syringe, and the cell was rinsed with DI water before using a new electrolyte. Electrochemical measurements were performed with a portable potentiostat (Ivium Technologies CompactStat). All potentials were referenced to the Ag/AgCl reference electrode.

### ACKNOWLEDGEMENTS

The authors acknowledge financial support from the European Research Council (ERC) under Horizon 2020 research and innovation programme PICOFORCE (Grant Agreement No. 883703), and from the EPSRC (Cambridge NanoDTC EP/L015978/1, EP/L027151/1, EP/X037770/1). S.M.S.-T. is supported by the University of Cambridge Harding

Distinguished Postgraduate Scholars Programme. S.M.S.-T., acknowledges further support from EPSRC Grant EP/L015889/1 for the EPSRC Centre for Doctoral Training in Sensor Technologies and Applications. R.A. acknowledges support from St. John's College Cambridge. M.N. is supported by a Gates Cambridge fellowship (OPP1144). B.d.N. acknowledges support from the Royal Society (URF\R1\211162) and the EPSRC (EP/Y008294/1).

## Supporting Information
Supporting Information is available from the journal website.

## Author Contributions:
E.W.W. and J.J.B. designed the experiments, which were carried out by E.W.W. S.M.S.-T., RA., M.N., J.W.B. and T.J. aided with substrate fabrication, and data interpretation. Data analysis was by E.W.W., B.d.N., and J.J.B., and all authors contributed to the manuscript drafting and editing.

## Conflict of interest:
The authors J.J.B., E.W.W., S.M.S.-T., R.A., and M.N. declare the following competing interests: filed patent, Surface-enhanced spectroscopy substrates, UK 2304765.7, 30/3/2023.